\documentclass{emulateapj}

\usepackage{graphicx}
\usepackage{epsfig}
\usepackage{natbib}  
\usepackage{epstopdf}
\def\simgt{\lower.5ex\hbox{$\; \buildrel > \over \sim \;$}}
\def\simlt{\lower.5ex\hbox{$\; \buildrel < \over \sim \;$}}

\newcommand{\msun}{\ensuremath{\, {M}_\odot}}
\newcommand{\Msun}{\ensuremath{\, {M}_\odot}}
\newcommand{\ocen}{$\omega$~Cen}
\newcommand{\tbce}{T$_{\rm bce}$}

\shorttitle{The O--Na (anti)correlation(s) in \ocen}
\shortauthors{D'Antona et al.}

\begin{document}

\title{The oxygen vs. sodium (anti)correlation(s) in \ocen}


\author{F. D'Antona }
\affil{INAF -- Osservatorio di Roma, via Frascati 33, I-00040 Monteporzio. Italy}
\email{francesca.dantona@oa-roma.inaf.it}

\author{A. D'Ercole}
\affil{INAF -- Osservatorio di Bologna, via Ranzani, 1, I-40127 Bologna. Italy}
\email{annibale.dercole@oabo.inaf.it}
\author{A.F. Marino}
\affil{Max-Planck-Institut f\"ur Astrophysik, Garching, Germany}
\email{ amarino@MPA-Garching.MPG.DE}

\author{A. P. Milone}
\affil{IAC-Instituto de Astrofisica de Canarias, \& Department of Astrophysics, University of La Laguna, V\'ia L\'actea s/n, E-38200 La Laguna, Tenerife, Canary Islands, Spain}
\email{milone@iac.es}

\author{P. Ventura}
\affil{INAF -- Osservatorio di Roma, via Frascati 33, I--00040 Montepozio, Italy}
\email{paolo.ventura@oa-roma.inaf.it}

\and

\author{E. Vesperini}
\affil{Department of Physics, Drexel University, Philadelphia, PA 19104, USA}
\email{vesperin@physics.drexel.edu}



\begin{abstract}

Recent exam of large samples of \ocen\ giants shows that it shares with mono-metallic globular clusters the presence of the sodium versus oxygen anticorrelation, within each subset of stars with iron content in the range --1.9$\simlt$[Fe/H]$\simlt$--1.3. These findings suggest that, while the second generation formation history in \ocen\ is  more complex than that of mono-metallic clusters, it shares some key steps with those simpler cluster.  In addition, the giants in the range --1.3$<$[Fe/H]$\simlt$--0.7 show  a {\it direct} O--Na correlation, at moderately low O, but Na  up to 20 times solar. These peculiar Na abundances are not shared by stars in other environments often assumed to undergo a similar chemical evolution, such as in the field of the Sagittarius dwarf galaxy. These O and Na abundances match well the yields of the massive asymptotic giant branch stars (AGB) in the same range of metallicity, suggesting that the stars at [Fe/H]$>$--1.3 in \ocen\  are likely to have formed directly {\it from the pure ejecta of massive AGBs of the same metallicities}. This is possible if the massive AGBs of [Fe/H]$>$--1.3 in the progenitor system evolve when all the pristine gas surrounding the cluster has been exhausted by the previous star formation events, or the proto--cluster interaction with the Galaxy caused the loss of a significant fraction of its mass, or of its dark matter halo, and the supernova ejecta have been able to clear the gas out of the system. The absence of dilution in the metal richer populations lends further support to a scenario of  the formation of second generation stars in cooling flows from massive AGB progenitors. We  suggest that the entire formation of \ocen\ took place in a few 10$^8$yr, and discuss the problem of a prompt formation of s--process elements.
\end{abstract}





\section{Introduction}
\label{sec:intro}
A  number of surprising observational findings of the latest dozen years suggest a completely new paradigm for the formation of Globular Cluster (GC)  stars. 
The abundances of stars in most Galactic GCs observed so far  show particular patterns, such as the oxygen vs. sodium and magnesium vs. aluminum anticorrelations, not shown by the halo stars \citep{gratton-ar}, both in giants and in scarcely evolved stars \citep[e.g.][]{gratton2001, ramirezcohen2002,gratton-ar, carretta2009a,carretta2009b}. 
The observed amount of chemically anomalous stars and the relative number of normal and anomalous stars  suggest a sequence of events  starting with the birth of a first stellar generation (FG) in a ``protocluster" much more massive than todays' cluster, followed by the birth of a second generation (SG) from gas having the same iron content, but including hot--CNO cycled matter ejected by the FG stars evolving during the first phases of the cluster life  \citep[for a summary, see, e.g. ][]{dercole2008}. 
Theoretical studies have identified the matter constituting the SG stars as the gaseous ejecta either by massive asymptotic giant branch (AGB) stars \citep[``AGB  scenario";][]{ventura2001, dantonacaloi2004, karakas2006} or by fast rotating massive stars \citep[``FRMS scenario";][]{prantzos2006,   meynet2006, decressin2007}. These studies investigated whether the observed physical properties of the Galactic GCs are consistently accounted for in the models. Problems are present in either scenario \citep[for a critical  discussion, see][]{dercole2010}, although some chemical signatures of the SG, such as the Lithium abundances \citep{dorazimarino2010,shen2010} or the magnesium depletion \citep{vd2009, vcd2011} favour the AGB scenario, in which the hot CNO processing occurs by ``hot bottom burning" (HBB) in the convective envelopes, that are recycled to the intra--cluster medium by wind mass loss. The AGB ejecta are likely to be converted into new stars, thanks both to their low ejection velocities and to the fact that they evolve after the end of the energetic Supernovae II epoch.

Most GCs show none or scarce iron spread among their stars \citep[e.g.][]{carretta2009ferro}.  A few clusters, however, must be born in a more complex ``protocluster" environment. It has been often suggested that \ocen, the most massive galactic GC, formed in a now dispersed dwarf galaxy \citep[e.g.][]{norris1997,rey2004, bekki-norris2006} and, possibly, has an evolutionary history similar to that of  M54, whose host galaxy, the Sagittarius dwarf,  is now merging into the Galaxy.   Both \ocen\ and M~54 show a large spread in iron. \ocen\ contains stars with iron contents spanning a range from [Fe/H]$\sim$--2 to [Fe/H]$\sim$--0.6 \citep{norris1996, suntzeff1996, lee1999, pancino2000, sollima2005a, johnson2008, johnson2010, marino2011}. 

The complexity of \ocen\ star formation episodes is also shown from  its multiple red giant and sub giant branches \citep{lee1999, pancino2000, sollima2005b}. 
\cite{bedin2004} found that the main sequence (MS) contains a well separated blue branch, probably then hiding a large overabundance of He \citep{norris2004}, as its stars are more iron-rich than stars on the redder side \citep{piotto2005}. A third, less populated MS (MSa) has been discovered by \cite{bedin2004} on the red side of the red MS and it has been associated to the most metal rich stars \citep[e.g.][]{villanova2007}. As well as blue MS stars, the colors and the metallicity of the MSa suggest that it is populated by He rich stars \citep{bellini2010}. 

The presence of large variations in the light elements abundances and s--process elements is well documented \citep[see, e.g.][]{paltoglou1989, ndc1995, smith2000}. \cite{ndc1995} show that \ocen\ giants also display huge aluminum variations, anticorrelated with oxygen and correlated with sodium, another typical pattern present in mono--metallic GCs. More recently, the spectral analysis of a sample of 840 giants by \cite{johnson2010} and 300 giants by \cite{marino2011} shows with more clarity a very interesting peculiarity. Although the presence of a wide range of iron contents points towards a chemical evolution similar to that of galactic environments,  the stars in each bin of iron content for [Fe/H]$<$--1.3 {\it also show the typical O--Na anticorrelation of GCs having a homogeneous iron content}. On the contrary, at [Fe/H]$>$--1.3 the O and Na abundances show a direct correlation, and the [Na/Fe] values are always much larger than the solar ratio.

The attempts to model the chemical evolution of \ocen, in the framework of a dwarf galaxy environment, have not been fully successful \citep[e.g.][]{romano2007,choi2008} especially for the difficulty in modeling the origin of the very helium-rich population(s). 
\begin{figure*}    
\plottwo{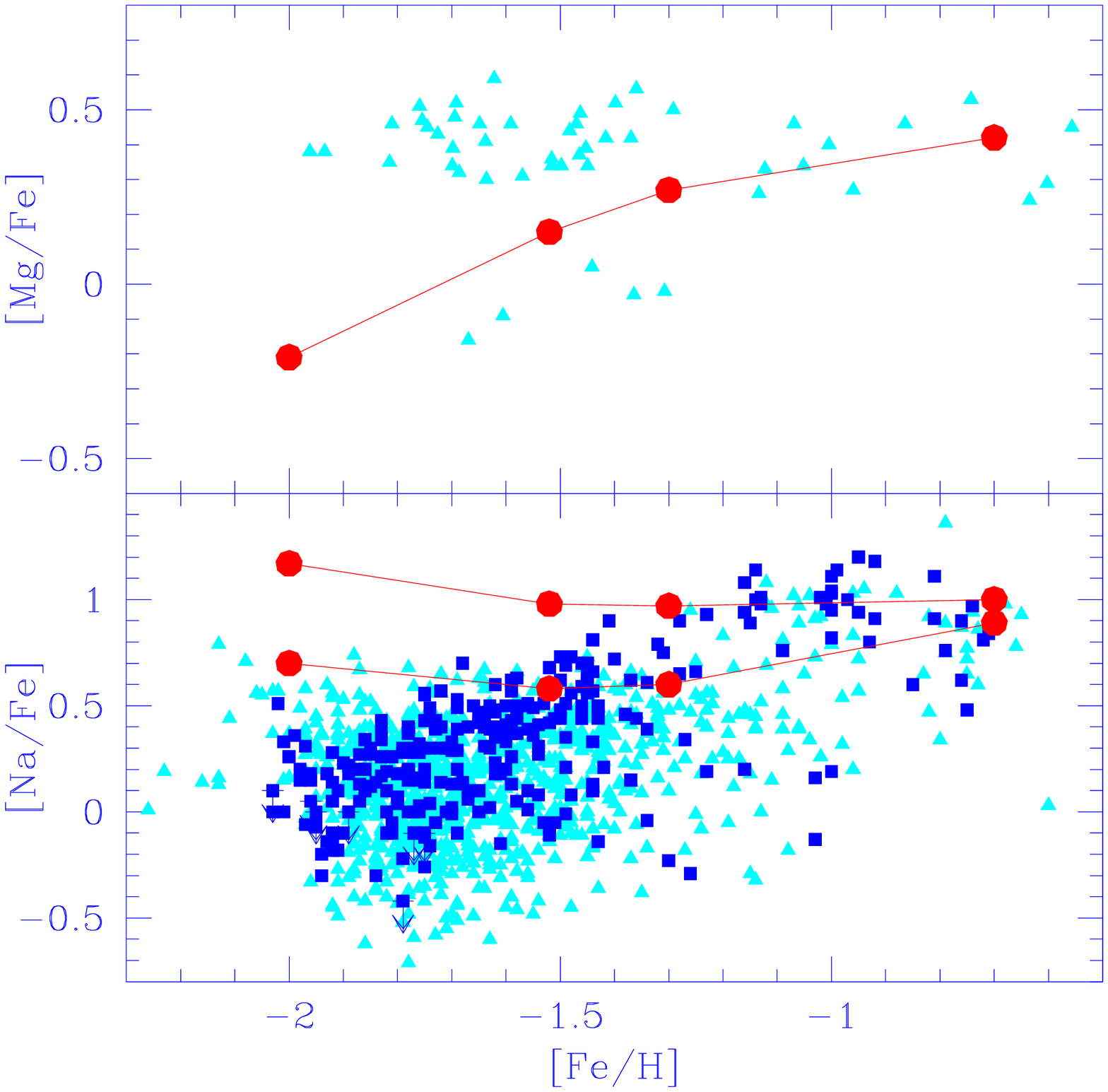}{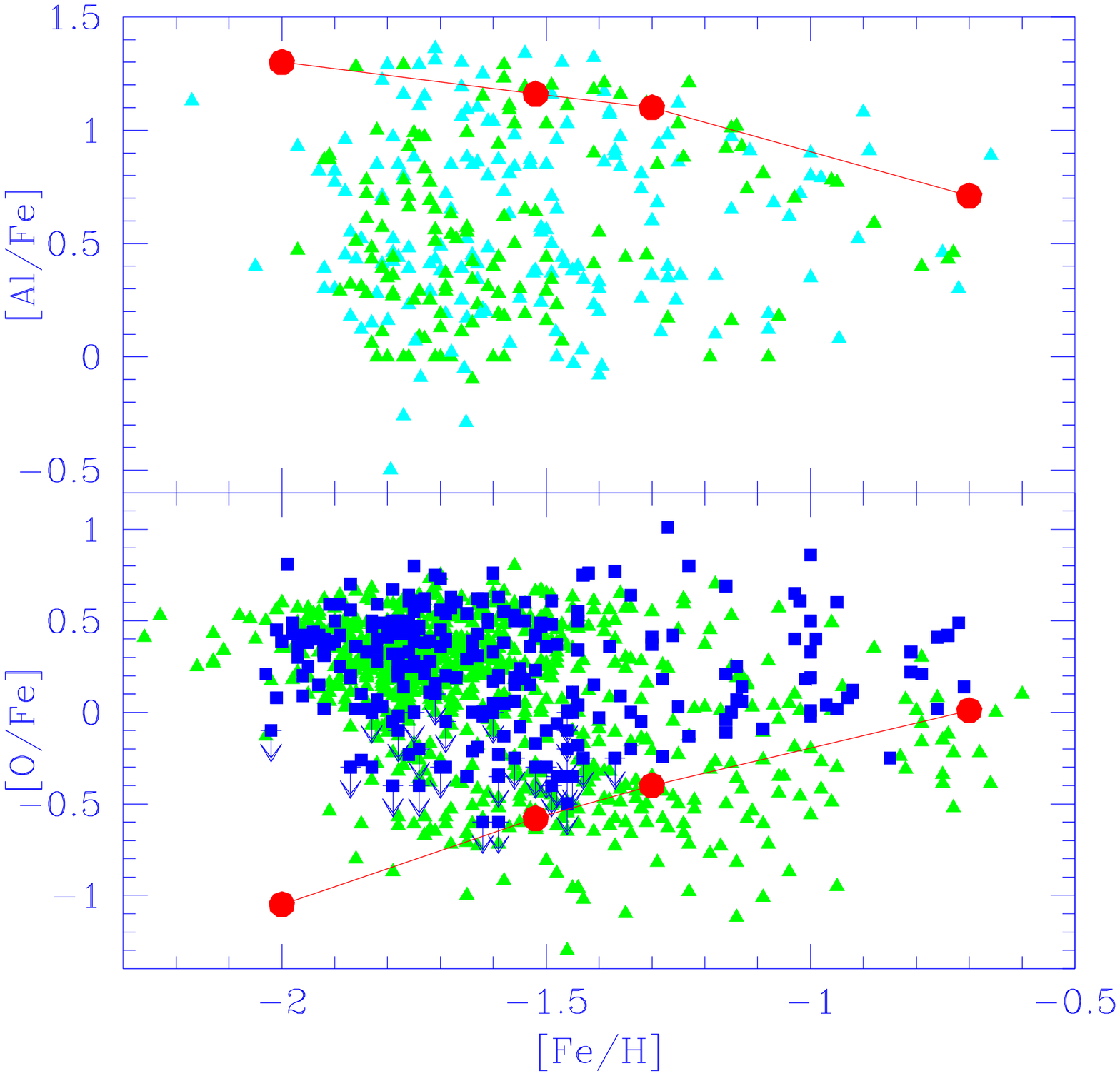}
\caption{Literature data for O, Na, Mg and Al as a function of the iron content [Fe/H]. 
The lower boundaries for Na and Al and the upper boundaries for O and Mg both follow closely to the trends predicted from SN II ejecta, weighted by a ``typical" initial mass function.
The top left panel shows the Mg data by Norris \& Da Costa (1995), the bottom left panel shows the Na data by Johnson \& Pilachowski (2010) (triangles) plus the data by Marino et al. (2011) (full squares). The top right panel shows the  Al data by Johnson et al. (2008, 2009) (cyan triangles) and Johnson \& Pilachowski (2010) (green triangles), and the bottom right panel shows the Johnson (2010) O data (triangles) plus the data by Marino et al. (2011) (full squares). The full big dots (red) in the panels represent the AGB yields as a function of [Fe/H] from \cite{vd2009}, and in particular the {\it minimum} Mg and O yields, and the {\it maximum} Al yield. For sodium, we plot two values: the yields for 5\msun\ (lower line) and 4.5\msun. }
\label{f1ab} 
\end{figure*}
The most recent modelling by \cite{romano2010} does not yet include the spectroscopic results concerning the O--Na anticorrelation {\it for each subset of iron content} we just discussed. 
In fact, a linear chemical evolution path for each element can not describe this complex situation. Their work has to make a series of complex  hypotheses concerning  winds that must preferentially eject some products of nuclear evolution, in order to reproduce some characteristics of the cluster stars. In particular, a timescale of a few Gyrs is needed to form the stellar generations with different iron content in this cluster, and this rules out the massive AGB scenario for the formation of the super--helium rich population(s); the formation of this population, according to \cite{romano2010} should then be attributed to the yields of FRMS, but see, e.g., \cite{renzini2008} against this possibility. Since our work is focused on the AGB scenario also for the formation of the very helium rich population(s), we will discuss later on (Sect.\ref{sprocess}) why this fails in the approach by \cite{romano2010}. 

We point out another ingredient of the \ocen\ puzzle: several studies have shown that the blue MS stars are more strongly concentrated in the cluster core than the red MS stars, and that this higher concentration holds for the most metal rich populations too \citep{suntzeff1996, norris1996, rey2004, bellini2009, johnson2010}. These observed differences in the spatial distribution of first- and second-generation stars are consistent with the predictions of the simulations presented in \citet{dercole2008}. Specifically, \citet{dercole2008} show that the SG formation in ``simple", mono--metallic GCs can occur in a cooling flow that gathers the massive AGB ejecta in the cluster core \citep[see also][for 3D simulations confirming the results of D'Ercole et al. 2008]{bekki2010}.

\cite{dercole2010} extended the dynamical model to predict the abundances in the SG and, in particular, how the O--Na and Mg--Al anticorrelations are built up. 

The remarkable O--Na--Fe abundance distribution described above, introduces a new complex ingredient in the evolution of \ocen\ and suggests that a new approach is needed to model the chemical evolution of this cluster.
We suggest that, despite the complexity of the abundance patterns in \ocen\ stars, the formation history of this cluster may be explained according to the key steps of the \cite{dercole2008} model, that also provides the key to understand how the stars with [Fe/H]$>$--1.3  may be directly born from the pure ejecta of massive AGB stars \citep[as already recognized by][]{johnson2009}. We also show that the cluster formation outlined in this work is in agreement with the abundances of other light elements (Mg and Al). It is also consistent with the spatial distribution of \ocen\ populations. The critical problem of the  formation timescale(s) of the multiple populations constituting the GC as we see it now is finally discussed in Sect.\ref{sprocess}, where we remark that the abundances of s--process elements as a function of metallicity may be uncorrectly biasing our views. 

\section{The p--capture data in \ocen}
\label{dilut}
While in this work we will focus mainly on the Na and O abundances in \ocen, and will use the data from  \cite{marino2011} for the sake of homogeneity, here we provide some more information on results   for Mg and Al, whose abundances constitute another powerful indicator of p--capture processing in the gas from which the SG stars formed. Fig. \ref{f1ab} shows the O, Na, Mg and Al abundances as a function of [Fe/H] from different data, and the lines corresponding to theoretical models. These latter will be discussed in Sect.~\ref{mgal}. Mg data are taken from \cite{ndc1995}, and Al data are from \cite{johnson2008, johnson2009, johnson2010}. The figure shows that Na and O data from \cite{johnson2010}  are qualitatively compatible with the  \cite{marino2011} data, although a trend to have lower Na and O values is evident, especially at [Fe/H]$>$--1.3. \cite{marino2011} show in their Fig.~6 that their own oxygen determinations can be larger than the \cite{johnson2008, johnson2009} and \cite{johnson2010} determinations, and the difference may reach 0.3~dex or more at low oxygen values. 
Similar systematic (although smaller) differences appear also in the sodium determinations. For this reason we decided not to combine the different datasets, and we mainly use our own data to put constraints on the O--Na (anti)correlation(s).

\begin{figure}    
\plotone{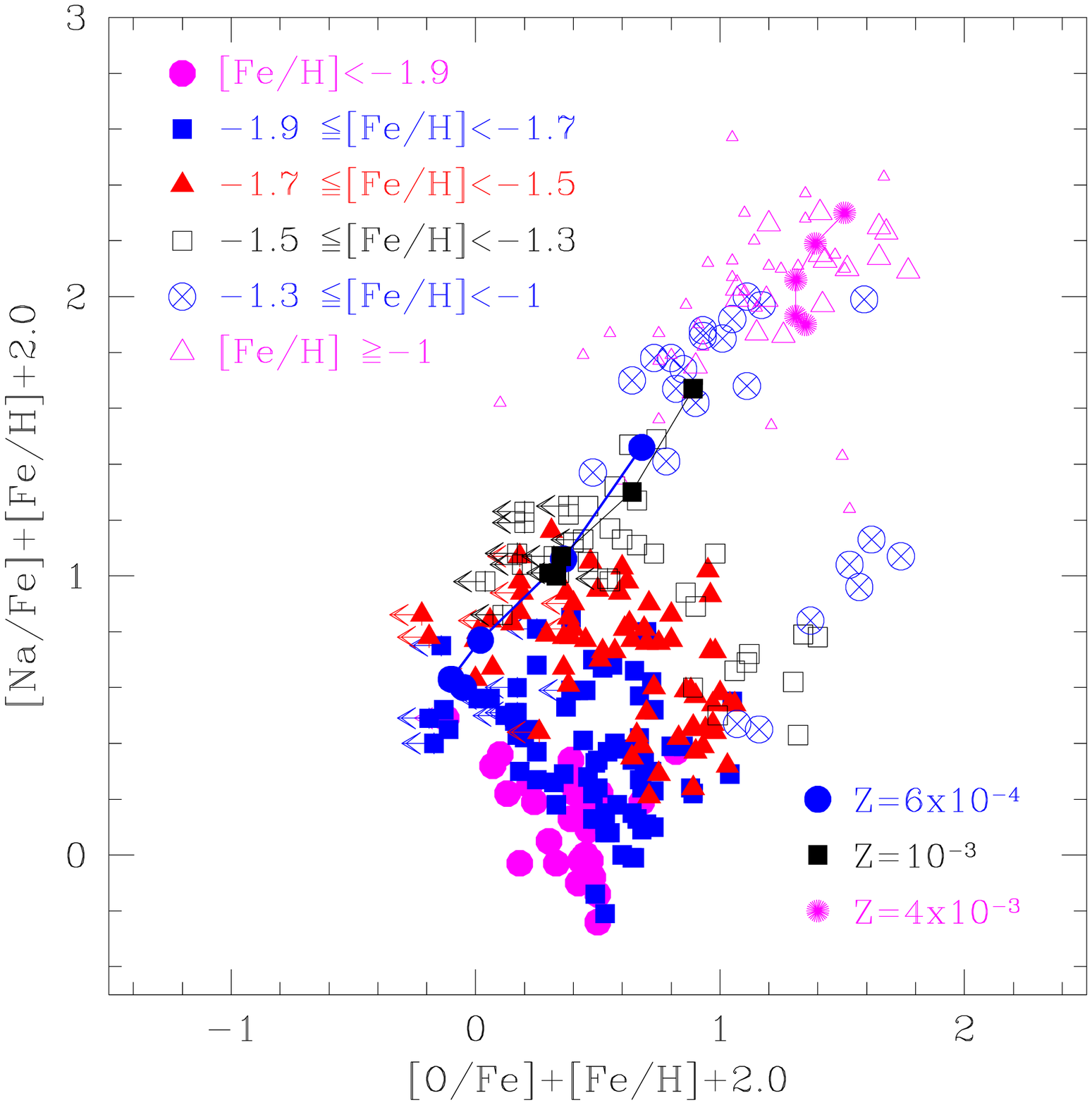}
\caption{The Na vs. O data by Marino et al. (2011) for the giants in \ocen\ are subdivided into different iron content groups, and plotted with the abundances are normalized to [Fe/H]=--2. The open triangles are the data for [Fe/H]$\geq$--1 (large triangles from \cite{marino2011}, small ones from \citet{johnson2010}).
The lines with symbols on the left are the yields from the models by Ventura and D'Antona (2009) for the labelled metallicities. Masses from top to bottom of each segment  are 4.5, 5, 5.5, 6\msun. The last point is  6.4\msun (Z=6$\times 10^{-4}$), 6.5\msun (Z=$\times 10^{-3}$) or 6.3\msun (Z=4$\times 10^{-3}$).}
\label{f2} 
\end{figure}

We show the O and Na abundances from the data by \cite{marino2011} in Fig.~\ref{f2}. We consider 6 bins of metallicity, keeping in mind that the errors on the abundance determination may easily shift some points from one bin to the adjacent one(s)  \citep{marino2011}. We plot [O/H] and [Na/H] instead of the values normalized to the iron abundance, scaling all the values to the lowest metallicity [Fe/H]=--2. In order to increase the  relatively few data at [Fe/H]$>$--1, we add the values by \cite{johnson2010} for these metallicities. These data increase the scatter, but, in this kind of plot, they do not change the qualitative behaviour of the metal rich sub--populations.  We point out here some interesting patterns emerging from this figure (see also the right panel of Fig.~\ref{f3}):  1) the lowest metallicity bin shows a scarce, if any, anticorrelation; 2) {\it each} intermediate metallicity bin in the range --1.9 \simlt [Fe/H]$<$--1.3  shows {\it independent} extended O--Na anticorrelations;  3) the highest metallicity bins (apart from a few stars in the range --1.3 \simlt [Fe/H] $<$ --1.1 located in the O--rich,  low--Na  region) show very clearly a radically different behaviour, a {\it direct} O--Na correlation, with oxygen  abundances slightly reduced  and very large sodium values, up to $\sim$20 times the solar ratio.

The huge sodium values of the most iron rich bins is a very atypical feature; we are not aware of  any other galactic environment in which such a feature is found. In particular, the field stars of the Sagittarius dwarf galaxy having comparable iron contents, have solar, or lower than solar, [Na/Fe] \citep{sbordone2007, carretta2010sag}. Thus the field of the Sagittarius formed in a very different context than the metal--rich population in \ocen, although other elemental abundances seem to suggest a similar chemical evolution \citep{carretta2010m54wcen}.  

In Fig.~\ref{f2} we also plot the yields of the massive AGB models (see Fig.~\ref{f3} for the yields down to 3\msun), scaled in the same way as the data. We limit the masses to the range 4.5$\leq$M/\Msun$\leq$6.5.  Both O and Na increase when the stellar mass decreases (see Fig.~\ref{f3} for the super--AGB behaviour). In fact, smaller masses have lower temperature at the HBB boundary of convection (\tbce), and only allow a milder ON and Na cycling.  Increasing the metallicity, the models define parallel and partially superimposed diagonal lines, that shift to larger O and Na values for larger metallicity. This is due to two concomitant reasons. 
The first reason is that, by increasing the metal content, the average content of each element increases. The sodium yield is a result of four different processes occurring in the HBB envelopes: 
1) second dredge up of the sodium produced in the stellar interior in the previous phases of evolution; 2) burning to sodium of the $^{22}$Ne achieved during the same process of second dredge up; 3) and burning during the HBB phase along the Ne--Na cycle. 4) In addition, in the relatively small masses (M$\le$4.5\msun), sodium is also produced by HBB of the $^{22}$Ne  dredged in the envelope at each third dredge up process. 
As sodium is produced from $^{20}$Ne and  $^{22}$Ne, and this latter is ultimately produced from $^{14}$N, the average production of $^{23}$Na, {\it for   similar HBB conditions} must increase with stellar metallicity.  For the same reason, the equilibrium value of O in the full CNO processing is larger at larger C, N and O abundances. 

The second reason is that the HBB efficiency decreases with increasing metallicity, as the higher opacities affect the envelope structure and allow smaller \tbce's, lower ON processing and Na depletion \citep{vd2009}. 
Notice that  the computation of the sodium yields  is a delicate issue, as they depend on uncertain cross sections, and also on the initial abundance of $^{20}$Ne. This problem is examined extensively by \cite{dercole2010}, who adopt Na abundances larger by 0.2--0.3dex with respect to those plotted in Fig.~\ref{f3}. Yields from other modelers \citep[e.g.][]{karakas2007} attain smaller oxygen depletion and larger sodium abundances for a fixed mass, due to the different assumptions made for the convective efficiency and mass loss. This issue is largely discussed in \cite{vd2005a, vd2005b}.

The theoretical yields seem to define a border limit to the abundances of the data, and the yields of the massive AGBs of the highest available metallicity Z=$4\times 10^{-3}$\ go through the data in the two bins of largest iron content. As the model yields reproduce the observed O--Na correlation for the high-metallicity population, we are led to suggest that the stars  at [Fe/H]$>$--1.3 {\it form from the pure ejecta of massive AGBs of the same  metallicity}, and this may represent a key ingredient to understand the key lines of chemical evolution in \ocen.  If we had adopted \cite{johnson2010} data, the same conclusion would have been reached, but only for [Fe/H]$>$--1, as their oxygen values for giants having  --1.3$<$[Fe/H]$<$--1 can reach much lower abundances, likely due to systematics both in [Fe/H] and   [O/Fe] between the two data-sets \citep[see Figs.~4--6 in][]{marino2011}. We now try to use this finding as a constraint for the formation model of the cluster.

\section{The O--Na plane: data points, theoretical yields and predictions of D'Ercole et al. model}
\label{theory}
\begin{figure*}    
\plottwo{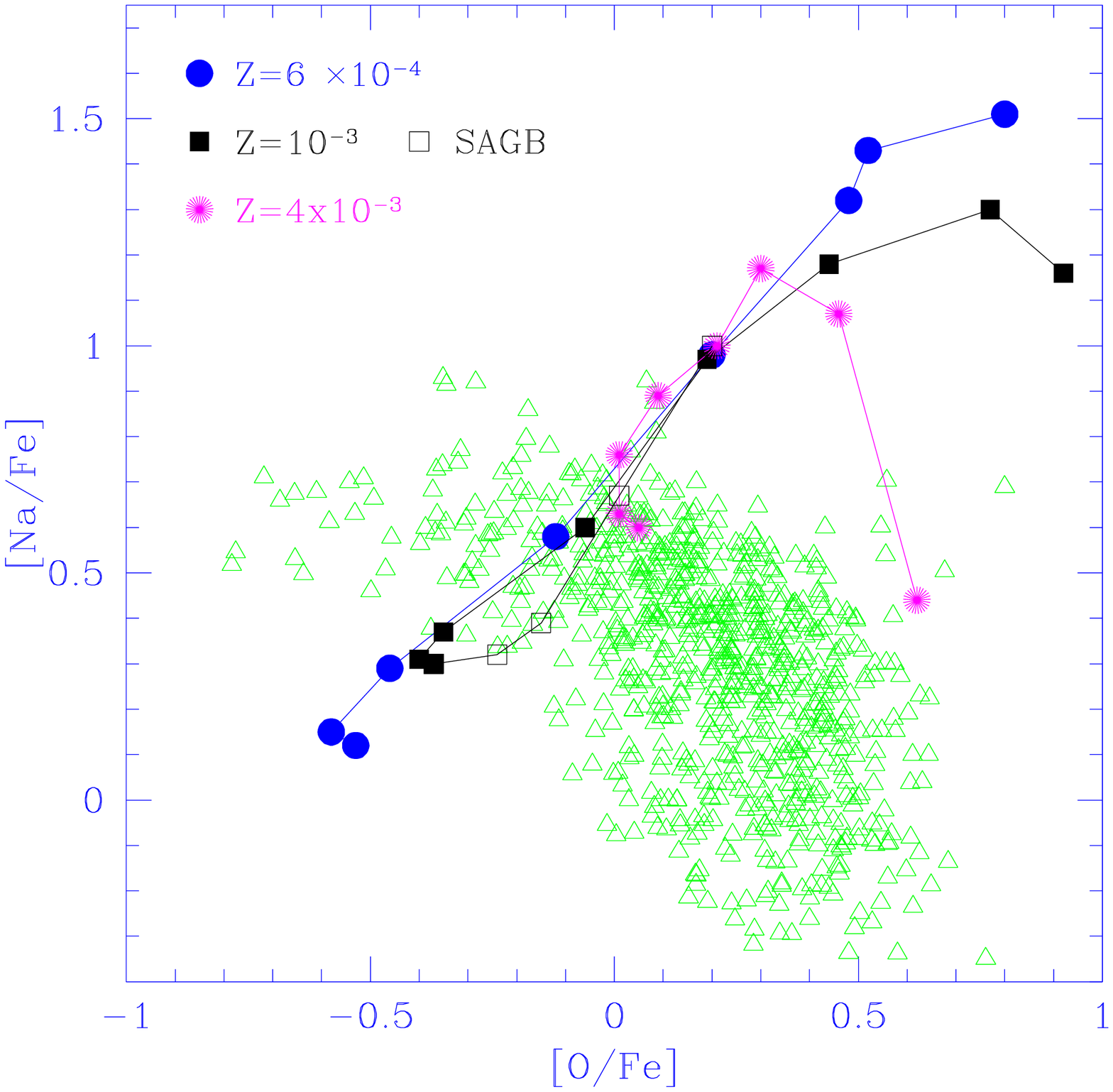}{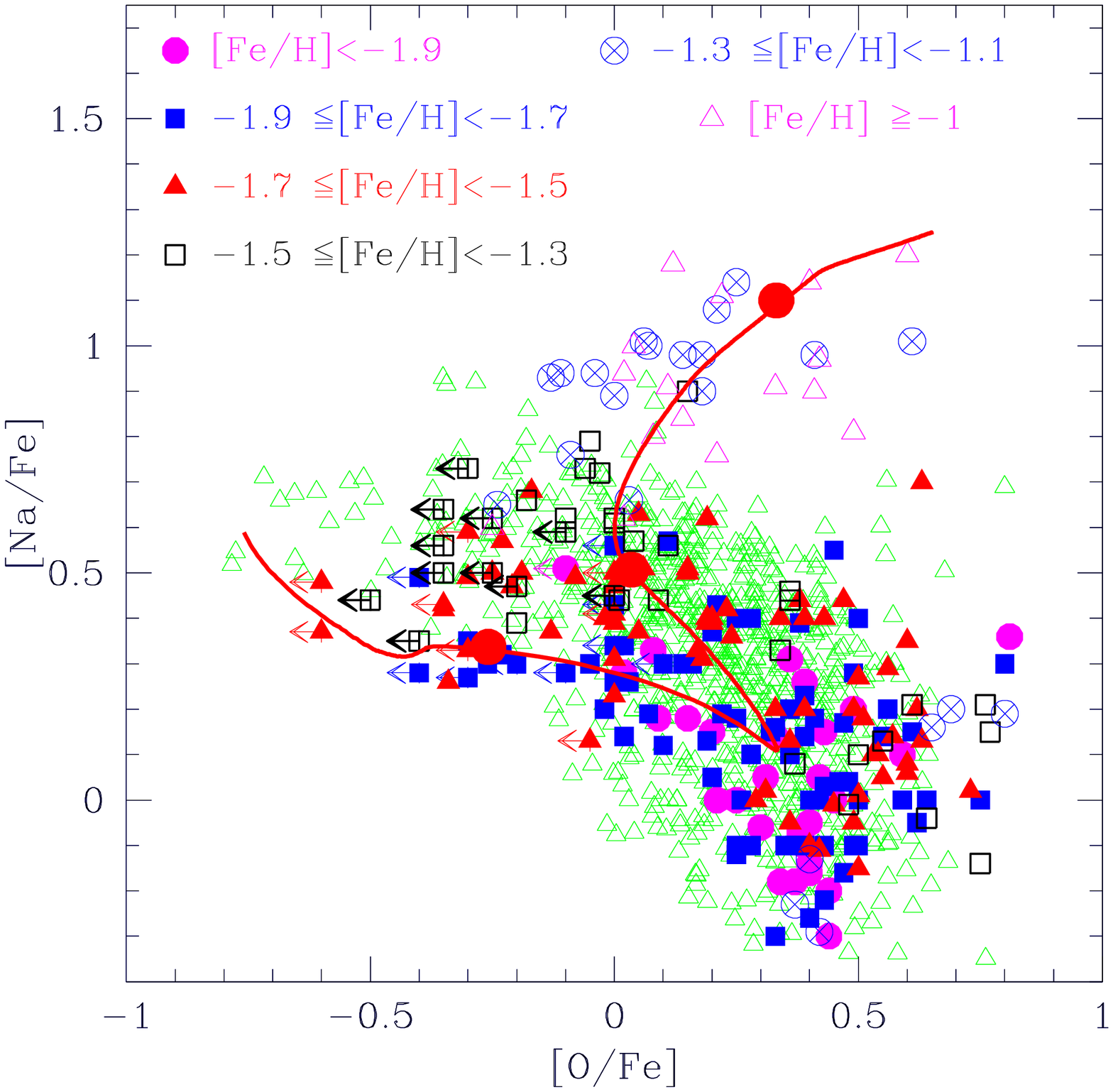}
\caption{Left panel: Na vs. O data for GCs in Carretta et al. (2009a) are shown as open triangles. Superimposed, we plot the yields for AGBs from Ventura \& D'Antona (2009). Metallicities are labelled. Masses from top to bottom of each segment  are 3, 3.5, 4, 4.5, 5, 5.5, 6\msun. The last point is  6.4\msun (Z=6$\times 10^{-4}$), 6.5\msun (Z=$\times 10^{-3}$) or 6.3\msun (Z=4$\times 10^{-3}$). The Z=10$^{-3}$\ yields are completed with the models by Ventura \& D'Antona (2011) for super--AGB models of 6.5, 7, 7.5 and 8\msun, that revert upwards to larger oxygen and sodium abundances.
Right panel: The Na vs. O data by Marino et al. (2011) for the giants in \ocen\ are plotted on the data of mono--metallic GCs shown in the left panel. The stars are subdivided into different iron content groups. The line represents the chemical pattern of the SG in a monometallic cluster with Z=10$^{-3}$, according to the standard model by \cite{dercole2008}, updated using the model yields of the left panel and the prescriptions described in the text. Dots label ages of  80, 100 and 150Myr from the formation of the FG. The line ends at 200Myr.
}
\label{f3} 
\end{figure*}
Before we discuss a possible modelling of the O--Na data in \ocen, it is propaedeutic to summarize the status of art of modelling the O--Na anticorrelation in single--metallicity GCs. The left panel of Fig.~\ref{f3} shows the data by \cite{carretta2009a} for giants in these clusters. Superimposed, we plot the whole oxygen and sodium yields of AGB stars from \cite{vd2009}.
As discussed in Sect.~\ref{dilut}, masses increase from top right to bottom left. For the Z=10$^{-3}$\ models, we also plot the results of the yield computation by   \citet{vd2010b} for the masses 6.5$\leq$M/\msun$\leq$8, that, after having ignited Carbon in conditions of semi--degeneracy, and formed an O--Ne degenerate core, follow the super--AGB phase. We see that these more massive progenitors do not change the location of the Na--O model correlation, but simply reverse it, and now the O and Na content increase for increasing mass. This is due to the huge mass loss rates of super--AGBs. So by increasing the initial mass from 3\Msun\ to the maximum mass that does not explode as supernova, the Na--O line first goes down, reaches a minimum value in the yields (at masses $\sim$6.5\msun), and then reverts back to larger O and Na. 

Three main features are evident from the comparison between yields and single-metallicity globular cluster data:
\begin{enumerate}
 \item the AGB models of different masses provide a {\it direct} correlation between sodium and oxygen abundance in their ejecta. The presence of the O--Na anticorrelation in the GC data is generally attributed to dilution of the hot--CNO processed gas of the stellar ejecta with ``pristine" gas \citep{bekki2007,dv2007,dercole2008, dercole2010}. Dilution is also needed when dealing with FRMS ejecta \citep{prantzos2006,decressin2007,lind2011}. In all clusters dilution seems to be present. Detailed  hydrodynamical models  able to explain the dynamics and the presence of pristine gas during the SG formation of all clusters are still lacking. This issue is discussed in detail by \cite{dercole2011}.
 \item the yields for masses M$\simlt$4.5--5\msun lie outside the plane region occupied by the O--Na observational values. This represents a powerful constraint of the models, showing that the winds of massive AGB, for some reason, can not contribute to form SG stars at the epoch in which these --or smaller-- masses begin to evolve ($>$130Myr).  A powerful similar constraint is imposed also by the observed constancy of the   sum of CNO abundances among the stars of many GCs (e.g. M92 \citep{pilachowski1988}, NGC 288 and NGC 362  \citep{dickens1991}, M3 and M13 \citep{smith1996}, M4  \citep{ivans1999}) The occurrence of numerous third dredge up events during the AGB evolution of masses M$<$4.5\Msun\  \citep{vd2009,dercole2010} increases the total CNO. 
  \item the yields do not represent well the {\it lowest} oxygen values that can be present in the data.  \cite{dercole2010} assumed semi empirical yields down to [O/Fe]$=-1$\  in the super--AGB masses, by adopting a `deep--mixing" suggestion. \cite{dv2007} attribute the very small oxygen abundance present in the giant stars of the most extreme SG to  deep extra--mixing acting in the progenitors having a very high helium  content. In these stars, the small molecular weight discontinuity during the  red giant branch evolution may not be able to preclude  deep mixing and lowering of oxygen in the envelope. Stars having the largest helium abundances would then populate the blue MS and show the strongest chemical anomalies (see later). The presence of [O/Fe] abundances smaller than those predicted by AGB models only among the giants, and not among the turnoff or subgiant stars so far examined \citep{carretta2006}, and in stars belonging to clusters also showing the signature of a very helium rich MS, such as NGC~2808 and \ocen, are a possible argument in favour of this interpretation. Obviously, the determination of a very low oxygen content ---similar to the one in the  most anomalous red giants--- in the atmosphere of blue MS stars in these clusters would falsify this suggestion \citep[][were not able to measure oxygen in their spectrum of the blue MS in NGC~2808.]{bragaglia2010}.
\end{enumerate} 
Let us discuss how the O--Na anticorrelation is introduced in \cite{dercole2008, dercole2010}. 
 In order to explain the data of NGC 2808, they propose a model in which the ``extreme" SG, corresponding to the most  oxygen-poor stars and to the stars populating the very helium rich blue MS, are born directly from the pure ejecta of the  super--AGBs of the FG. In fact, after the SN II epoch, that has cleared the cluster from the residual gas, a cooling flow sets in and brings to the core the low velocity winds of the super--AGBs, so that stars can form directly from their very  helium rich and hot--CNO processed material. Then a phase of mixing of the massive AGB ejecta with pristine gas   (reaccreted in the cluster core) follows, giving origin to the ``intermediate" SG stars, with milder oxygen depletion. The SG formation stops due to the onset of the Type Ia  SN epoch before  the pristine gas is exhausted. In the right panel of Fig.~\ref{f3}, the slanted V--shaped line shows an example of the O--Na patterns predicted by the \cite{dercole2010} model, based on the whole set of yields for masses from 3 to 8\msun\ shown in the left panel, and extended for 200Myr after the FG formation (D'Ercole et al. 2011, in preparation).
 Time along the line proceeds from the left side to down right, and then climbs up towards the upper right side. The ``extreme" SG is the left side of the V--shaped line, up to the point at $t$=80~Myr. The ``intermediate" SG stars follow the line from 80 to 100~Myr.  If star formation proceeds further, the subsequent SG would follow the right side of the V, from $t$=100~Myr to the end, when stars form directly from the ejecta of the evolving AGBs, so that the simulation follows the {\it direct} correlation O--Na shown in the left panel, but this side of the curve has no observational counterpart  in ``normal" GCs. The simulation stops at 2$\times 10^8$yr, when the evolving AGB mass is  $\sim$3.7\msun. The model is plotted on the [O/Fe], [Na/Fe] data for \ocen\ from \cite{marino2011}, and we see that data points are present also on the upper side of the simulation curve, but they have larger metallicity than stars showing the anticorrelation(s). This suggestive comparison requires some deeper understanding.

\section{Interpretation of the \ocen\ data in terms of D'Ercole et al. model}
We discussed in the previous Sect.~\ref{theory} how the O--Na anticorrelation in the GC data is attributed to dilution of the hot--CNO processed gas of the stellar ejecta with ``pristine" gas.  Where does this pristine gas come from? \cite{dercole2008} suggest that it may survive in a torus that collapses back on the cluster after the SN II epoch. If the system is much more massive than a typical proto--cluster as in the case of \ocen\ (possibly including  a dark matter halo) the SN II ejecta (and type Ia SN ejecta as well) can not be lost outside the potential well of the protocluster.  The 3D hydro simulations by \cite{marcolini2006} show in fact that the collapse back includes the matter enriched by the SN II ejecta. Their model was first developed to be compared with Draco's dwarf galaxy data. Afterwards, they attempted to model directly the case of \ocen\ \citep{marcolini2007}. Star formation in the cluster region stops when SN II explode, and a successive burst occurs when the interstellar medium (enriched by SN ejecta) flows back down the cluster potential well. Thus the gas content in the cluster center has an oscillating temporal profile whose period is given by the time-interval between two successive starbursts. Each starburst provides star formation with increasing metallicity.  If we add to this model the possible role of the winds of the AGBs evolving in the progenitor systems, these may account for star formation of the very helium rich stars when the cluster core is still devoid of matter (before each collapse back of the ISM), and for the formation of stars from AGB matter diluted with the ISM enriched by the SN II ejecta, similar to the events in the one--burst only model by \cite{dercole2010}. Suppose now that, after a last burst of star formation, there is no longer gas  that can collapse back from the outskirts of the cluster. Nevertheless, there can be further  episodes of star formation thanks to the cooling flows of the AGB winds. Whatever is the detailed mechanism by which the O--Na anticorrelation(s) are built in \ocen\ for [Fe/H]$<$--1.3, the O--Na correlation we see at larger metallicities is telling us that  the ISM in the outskirt of the cluster must be so reduced that only the AGB winds take part in the cooling flow and that these winds belong to the massive AGBs of the more metal rich populations. This interpretation only requires that these AGBs are delayed until there is no longer any pristine gas available
to collapse back in the core of \ocen.  A possible hypothesis is that the gravitational interaction of the \ocen\ progenitor with the Galaxy \citep{dinescu1999} caused the loss of a significant fraction of its mass, or of the dark matter halo, so that the last SN II burst, or the type Ia SN explosions could clean completely the central region and the whole proto--cluster from its residual gas.

Following this scenario, we now can re-interpret the path of the simulation shown in Fig.~\ref{f3} and explain why it resembles the shape of the \ocen\ data in the Na--O plane. In a mono--metallic cluster, the stars on the upper branch would be those formed by {\it smaller mass AGB stars ejecta} (M$\simlt$4.5\msun), evolving at a later time, when the diluting gas is over.  In \ocen, the high Na abundances belong to the stars in the most metal rich bins, and correspond to the yields of (again) more massive AGB stars for those metallicities (Fig.~\ref{f2}). The only requirements to explain why the most iron-rich giants in \ocen\ have large sodium and moderately low oxygen are: 1) these stars are formed in a cooling flow {\it directly} from the ejecta of the massive AGBs having the same metallicities; 2) the AGBs that provide the ejecta evolve at the time when all the gas surrounding the cluster has been already exhausted. 
 
\subsection{Consistency of the scenario: the magnesium and aluminum abundances}
\label{mgal}
We have so far focused our attention on the Na--O (anti)correlation(s) to outline a possible scenario for the formation of the multiple populations in \ocen\ because of the wealth of data available for the abundances of these two elements.

In this section, we  show that existing data on other light elements lend further support to the proposed scenario.
 In Fig.~\ref{f1ab} the top panels show  the Mg and Al data as a function of [Fe/H]. 
 %
 We plot on the data the {\it minimum} Mg yield and {\it maximum} Al yields from the \cite{vd2009} models. These values then represent the maximum degree of Mg burning and Al production in massive AGBs of each metallicity. The Mg data by \cite{ndc1995} show a remarkable constancy of magnesium for all metal contents, but the few low Mg points are at iron content for which models predict Mg depletion. 
 Quantitatively, the models are a factor $\sim$2 off the data, but see \cite{vcd2011} for a discussion of the dependence of the Mg depletion in AGB and super--AGBs on the input physics of the models. It is important to point out that {\it no Mg depletion} is predicted at the highest metallicities, for which, according to our model, the SG formation occurred directly from the pure AGB ejecta.
The upper boundary of Al data is in good agreement with the model yields. In particular, a smaller Al enhancement is predicted at high metallicity than at the lower ones, as observed. The Al data are also consistent with the hypothesis that some dilution of AGB ejecta with pristine gas occurs up to [Fe/H]$\simlt$--1.3, while either a dichotomy in the Al values is found at   --1.3$\simlt$[Fe/H]$\simlt$--1, or values compatible with pure ejecta at [Fe/H]$>$--1.
  
\section{The timescale of formation of the \ocen\ populations}
\label{sprocess}
We do not attach a fixed timescale to the model presented to account for the high sodium of the metal rich populations in \ocen. Its main requirement is that the time delay for the formation of the massive AGB progenitors in these iron bins must be of the order of the time required for the preceding formation events to exhaust the galactic gas in the neighborhood of the cluster, so that this gas does not collapse back to the cluster core after the last SN II epoch. We think that, in order to satisfy these requirements, the timescale for the formation of all the populations in \ocen\ must not exceed a few 10$^8$yr. 

Notice that the upper boundary for [O/Fe] (see the right panel of Fig. \ref{f3}) increases with increasing metallicity, in spite of the presence of the many O--poor giants defining the O--Na anticorrelation(s). Standard galactic chemical evolution  shows that [O/Fe] remains $>0$\ until the iron is produced by SN II, and shifts to the solar ratio only when Type Ia SN, producing much more iron and no oxygen, become the main source of metals \citep{matteuccigreggio1986}. Thus the iron chemical  evolution has been dominated by Type II SN, suggesting that the initial chemical evolution inside the progenitor system occurred on a {\it fast}  timescale. 
As also the other $\alpha$--elements seem to go towards the solar ratio only for the most metal rich population, corresponding to the MSa and RGBa \citep{pancino2002, origlia2003}, the timescale for the formation of the last, metal richest, population in \ocen\ must be in fact very close to the timescale for the onset of SN Ia explosions. Although this timescale is uncertain, its value is observationally estimated to be in the range 40--300 Myr \citep{mare01,mapa06,raskin2009}, and certainly not longer than 500Myr \citep[e.g.][]{galyam2004, sullivan2006}. Simulations adopting the double--degenerate model for this event show that the peak of the distribution is reached after only 250Myr \citep{tornambe1989}. It is difficult to see how this epoch can be delayed by more than a few 10$^8$\ yr in \ocen. In the  \cite{marcolini2006} and in other similar models, the presence of a dark matter halo prevents the clearing of gas by type Ia SN, so some other process must have precluded further star formation in the proto--\ocen\ (and in spheroidal dwarf galaxies as well), e.g. the gravitational interaction with the Galaxy.  

As the iron production is delayed in time and massive AGBs of different metallicity evolve at different epochs, also less massive (M$<$4.5\Msun) AGBs of the lower metallicity bins would contribute to the cooling flow. These may be, e.g., some of the stars in the bin --1.3$\leq$[Fe/H]$\leq$--1). 
To understand the role of smaller AGB masses of lower metallicities,
notice that the metal content of the massive AGB of high metallicity dominates over the anomalies of these latter stars (see the discussion of the O--Na yields shown in Fig.~\ref{f2}). As a net result, the [O/Fe] and [Na/Fe] of the gas may be not affected too much by the yields of the less massive stars  and the latter contribution to the average abundances remains a small perturbation. Of course, a full model is needed to quantify the extent of this problem.

HR diagram dating results in age differences from less than 2~Gyr \citep[e.g.][]{lee2005, sollima2005a} to 5~Gyr or more \citep{villanova2007} in recent literature. The influence of different helium content and/or of the total C+N+O content on the isochrones turnoff level may easily change these age determinations \citep{dantona2009, pietrinferni2009} and bring them to much closer values. Joo, S. \& Lee, Y.-W. (2011, in preparation) have indeed shown that it is possible that all the populations are coeval, by taking into account the combined effect of helium and metallicity.  The result is valid within the errors of $\pm$0.3~Gyr that affect the determination of relative ages from the turnoffs' location. This requires  that the MS-a is indeed helium rich, as suggested also by \cite{bellini2010}.

We address now the most important clue, that we consider a red herring, for the timescale of formation of this peculiar cluster, namely the evidence of a very fast and large growth, by more than a decade, in the abundances of Lanthanum and Barium \citep{johnson2010,marino2011} in the very small range of metallicity between [Fe/H]$\sim$--1.9 and [Fe/H]$\sim$--1.5.  This very dramatic s-process increase with iron, at such a low metallicity, is not found in other galactic environments  \citep{tolstoy2009}, as shown in Figure 17 of \cite{johnson2010}. \cite{marino2011} suggest a rapid production of these elements. The r-process is excluded, as it would also affect Europium as well, while the [Eu/Fe] ratio does not increase in the same metallicity interval \citep{smith2000, johnson2010}.  Another plausible production site are the cores of massive stars \citep[weak s--process, e.g.][and references therein]{raiteri1993}, but this chain should mainly increase the lighter nuclei in the s--chain, like $^{58}$Fe, $^{63}$Cu, or $^{65}$Cu \citep[see, e.g.][]{pumo2010}, while Cu in the metal rich stars of \ocen\  remains at low abundances \citep{cunha2002,pancino2002}. Further non--standard models for the production of these s--process elements may occur in rotating massive stars \citep{pignatari2008}. 

Nevertheless, the best acknowledged production site of these s--process elements  are the lower mass AGB stars (M$<$3\msun) \citep{busso1999} that evolve on timescales of several hundreds of million years or on gigayears timescales, and {\it these stars have been always considered the obvious responsible for this s--process production}. Of course, if the --say-- 3\Msun\ are responsible for the s--process production, this happens at a time at which all the possible producers of helium and of the O--Na anticorrelations (the stars of M$\simgt$5\msun !) have already evolved. We remark again that the 3\Msun\ stars would produce ingent amounts  of sodium, but do not deplete oxygen 
nor produce helium. It may well be that these masses contribute to increase the s--process abundances when the [s/Fe] value flattens (at [Fe/H]$\simgt -1.5$), but it is difficult to envision how they can contribute to the extreme s--process increase that occurs at --1.5$>$[Fe/H]$>$--1.9, a range of metallicities that show independent, and even extreme, O--Na anticorrelations (see the blue squares and the red triangles of Fig.~\ref{f2}) and where there is no correlation of the s--process abundances with, e.g., the oxygen depletion. So the problem is that the timescale to build up the O--Na anticorrelation(s) is much shorter than the timescale needed for standard s--process production sites. We do not see an easy way out of this problem, and in fact this is the reason why \cite{romano2010} had to exclude the AGB contribution for the formation of the helium rich population, and resort to the contribution of massive stars. 

We suggest that, in the same way as the O--Na anticorrelation occurs only in GCs, and is due to the peculiar chemical evolution provoked by the collection of winds from a restricted range of initial stellar masses, we may be facing a process that has not yet been explored in stellar models. A not well discussed site of s--nucleosynthesis occurs in the carbon burning shells of the tail of lower mass progenitors of SNII  \citep[e.g.][]{the2007}, and their contribution may peculiarly become apparent mainly in the evolution of the progenitor systems of clusters like \ocen, that are unable to get rid of the SN II ejecta, but still give birth to a cluster in which cooling flows play a dominant role. We remark that the only similar trend of increase in the s--process abundance, in the same range of metallicities, is found in M~22 \citep{marino2009,dacostamarino2010}, although this cluster completely lacks the most metal rich populations ([Fe/H]$>$--1.5) present in \ocen. Even the smaller cluster NGC~1851 shows an s--process increase by a factor 2--3 \citep{yong2008}, and this may be correlated to a possible, still discussed, small 
increase in the iron content \citep{carretta2010.1851,villanova1851.2010}. 
It would be interesting to know whether the cluster NGC~2419, that harbors two populations with a small variation in Calcium content \citep{cohen2010}, and a population with very high helium \citep{dicriscienzo2011} also shows a similar s--process dichotomy. 
It looks very difficult that these very different clusters managed to evolve in such a similar way, if there is not an unifying model for the s--process production, linked mainly to stars that evolve into SN II. Further work is required to clarify this issue.

\section{Conclusions}

We have examined the patterns of the  O--Na--Fe data by \cite{marino2011} for the giants in \ocen. Most stars at the largest iron contents ([Fe/H]$>$--1.3) are all  very sodium-rich and moderately oxygen depleted, a peculiarity not predicted in standard galactic evolution, and not found in other environments apparently similar, like in the metal rich field stars of the Sagittarius dwarf galaxy. Furthermore, the metal-rich stars populating the MS--a might be helium-rich \citep{bellini2010}. We have shown that these peculiar abundances (including high helium content) match those of the ejecta of the massive AGB stars of comparable metallicities \citep{vd2009}. 

We attribute the presence of a direct O--Na correlation for the stars at [Fe/H]$>$--1.3  to formation, in a cooling flow, from the ejecta of massive AGB stars of the same metallicity, that evolve in the cluster when the gas in the surroundings has been all exhausted by previous episodes of star formation. These previous events have formed the populations with [Fe/H]$<$--1.3, from the ejecta of AGBs, diluted with the recollapsed gas, metal enriched by the SN II preceding burst. This process forms the typical Na--O anticorrelation patterns. 
The reason why massive, metal rich AGBs are available for this phase of star formation is that, with respect to the AGBs at smaller iron content, they evolve with the time delay that has been necessary to form the metal-rich components in the progenitor system. 
Consequently, \ocen\ differs from mono-metallic clusters because its
initial mass  must have been even larger (possibly including a dark matter
halo) than that of a typical cluster.

Consequently, \ocen\ differs from mono--metallic clusters because its initial mass must have been even larger (possibly including a dark matter halo) than the initial mass of a typical cluster. This led to an evolution in which SN II ejecta could not be expelled from the cluster, but were incorporated in successive bursts of star formation, as described by \cite{marcolini2006}. 
The existence of helium-rich populations and the concentrated spatial distribution of the helium-rich and of the most metal-rich populations  \citep{suntzeff1996, norris1996, rey2004, bellini2009, johnson2010}, however, shows that also in this cluster the gas from which second generation stars form collects in a cooling flow in the cluster central regions  as described in the simulations by D'Ercole et al. (2008) for  standard GCs.
 A further reason, to think that these latter stars were formed in cooling flows, is the O--Na composition of their stars, that follows the direct O--Na correlation predicted by massive AGB stellar models. We showed that the formation of the \ocen\ multiple populations seems to follow the basic steps of the \cite{dercole2010} model for mono--metallic GCs. The O--Na patterns, and the Mg and Al abundances in \ocen\ then contribute to validate the AGB pollution model for the formation of multiple populations in GCs.

An important and still unsolved issue in the formation of \ocen\ multiple populations is which are the stars responsible for the fast increase in the s--process abundances, that occurs at the epoch in which the metallicity raises from $\sim$--1.9 to $\sim$--1.5. 
We discussed the common view of s--elements production in low mass AGB stars, and concluded that it is in contrast with the other strong indicators of formation of the GC populations in AGB--dominated cooling flows. We suggest that the s--process patterns are a red herring, and propose that their evolution is due to a process not yet identified or well explored in stellar models, such as,  e.g., the nucleosynthesis in the carbon burning shells of the tail of lower mass progenitors of SNII. This (postulated) nucleosynthesis site becomes dominant in the chemistry of clusters showing a spread in metallicity, in the same way as the O--Na anticorrelation, due to the peculiar chemical evolution provoked by the collection of winds from a restricted range of initial stellar masses, dominates the chemistry of GC stars. Further work is required to clarify this suggestion. For the same reasons, we suggest that the whole time of formation of \ocen\ is at most a few 10$^8$yr.


\acknowledgments
F.D., A.D. and P.V. have been supported by the PRIN-INAF 2009 grant  ``Formation and Early Evolution of Massive Star Cluster". E.V. was supported in part by the grant NASA-NNX10AD86G.
We thank Y.W. Lee for discussion on the problem of the age spread in \ocen, and A. Chieffi, M.L. Pumo and R.G. Gratton for discussion on the stellar sites for the s--process nucleosynthesis. We thank the referee for a constructive report.
We are grateful to A. Renzini for a critical reading of a first version of this manuscript.
This work is dedicated to F.D.'s grandson Filippo and his battle against leukemia.






\clearpage
\clearpage

\clearpage

\end{document}